# Long-range miniature ceramic RFID tags

Dmitry Dobrykh, Ildar Yusupov, Sergey Krasikov, Anna Mikhailovskaya, Diana Shakirova, Andrey Bogdanov, Alexey Slobozhanyuk, and Dmitry Filonov, and Pavel Ginzburg

*Abstract*— Radio frequency identification (RFID) is a mature technology, which allows performing contactless data readout via wireless communication links. While communication protocols in this field are subject to regulations, there is a room of opportunities to improve the hardware realization of antennas devices, which support the technology. In particular, readout range extension and miniaturization of passive RFID tags is an important objective with far going impact on retail, security, IoT, and many others. Here we introduce a new concept of high-permittivity ceramic tag, which relies on different physical principles. Instead of using conduction currents in metallic wires to drive electronic chips and radiate electromagnetic waves, high permittivity components rely on an efficient excitation of displacement currents. Those are efficiently converted to conduction currents, powering a memory chip. The practical aspect of this approach is improved robustness to environmental fluctuations, footprint reduction, and readout range extension. In particular, our high permittivity ceramic ($\varepsilon \sim 100$) elements have demonstrated a 25% reading range improvement in respect to commercial tags. In case, when state of the art readers and RFID chips are used, the readout distances of the developed ceramic tags are approaching 23 m and could be further extended with improved matching circuits.

## I. INTRODUCTION

Radio frequency identification (RFID) is a widely used technique for noncontact data readout via a wireless communication channel [1]. RFID elements are widely employed in warehouses logistics, billing systems, as biometric identifiers and in many other applications. Mature technology allows introducing RFID tags almost everywhere and tailors their electromagnetic design per application. The most commonly used architecture is based on passive ultra-high frequency (UHF) RFID tags and active interrogating readers, which establish connections and process time-modulated backscattered signals. While the communication channel is a subject to international regulations (e.g. EPCGEN2 standard), tag's and reader's antennae architectures are open to extensive research and optimization on pathways to improve performances, demanded by specific applications. For example, special tags are needed for labeling items with a metal surface. The later acts as a ground plane, which can quench a standard dipole-like RFID antenna, if additional efforts are not applied [2]. Ceramic materials are also used in cases, where tags should reliably operate under harsh conditions, including chemically aggressive environments and extreme temperatures. On the other hand, the textile industry requires integrating low-weight flexible miniature tags, protected against repeatable exposure to water [3]. There are dozens of other tailor-designed examples of RFID tags. Our objective, however, is to develop a new strategy to tag miniaturization and reading range extension without a relation to a specific application. For the sake of completeness, it is worth mentioning other types of tags – battery-assisted, active, and a rather new approach of chipless tags [4]. All of those, however, are outside of the scope here.

Typical architectures of RFID tags are based on folded metal strips with a chip, plugged in a gap. The operation of metallic resonators is based on phase retardation effects and, hence, those structures cannot be significantly smaller than a fraction of a wavelength. While quite a few techniques on size reduction do exist, antenna footprints cannot go significantly below the wavelength without a major bandwidth reduction. In theory, the overall radar cross-section (RCS) of a subwavelength structure does not depend on its overall size, if a proper resonant load is chosen. In practice, however, lumped element losses lead to significant performance degradations, making this approach to be less attractive in applications, where operation at moderately low signal-to-noise ratios is required. Extensive folding of metal strips can replace lumped elements and it is quite an attractive strategy for reducing internal losses. For example, meandered dipole antennas [5] are widely used in many RFID applications. Another example is the fractal dipole antenna [6]. Fractal geometries have self-similarity and space-filling nature and, being applied to antenna design, can provide a broadband operation at reduced spatial extend. Various methods are used to incorporate ground plane into designs. For example, patches, implemented on high-dielectric substrate materials and electromagnetic band-gaps are used to bring RFID tag antennae to the proximity of a metallic backplane [7], [8]. High permittivity substrates are also widely used for reducing tag's size [9]. Different techniques and types of materials have been used to develop electromagnetically contrast and mechanically flexible substrates for RFID antennae [10], including those, which allow placing tags on metal surfaces [11]. Metamaterials and compact volumetric geometries is another promising yet fully tested strategy for size reduction, e.g. [12],[13].

Another promising strategy in antenna size reduction is based on employing high-index materials, since the free space wavelength shrinks by the refractive index in all three dimensions. Dielectric resonant antennas (DRAs) [14] utilize this advantage towards demonstrating improved performances. In particular, DRAs usually have enhanced operational bandwidth and demonstrate robust performances in high-power applications. Preferable materials for DRA implementation are high-quality ceramics [15]. The permittivity of those composite materials can approach several thousand, opening a room of opportunities for electromagnetic design and miniaturization.

Here we investigate capabilities of miniature high permittivity dielectric resonators as RFID antenna elements. The general concept of the ceramic-based RFID tag appears in Fig. 1. The resonator is a ceramic cylinder, which supports localized magnetic mode, bounded to the structure. A reader device interrogates the tag and excites this magnetic mode. A miniature non-resonant metal split ring with a standard RFID chip, soldered in the gap, is placed on top of the dielectric cylinder. Near-field coupling converts displacement currents of the resonators into conduction current flow in the ring, initiating the operation of the chip. A backscattered modulated signal undertakes the reverse path – from the conduction current to the far-field radiation. The magnetic nature of the resonator's mode ensures an efficient coupling with the ring via the inductive-like coupling.

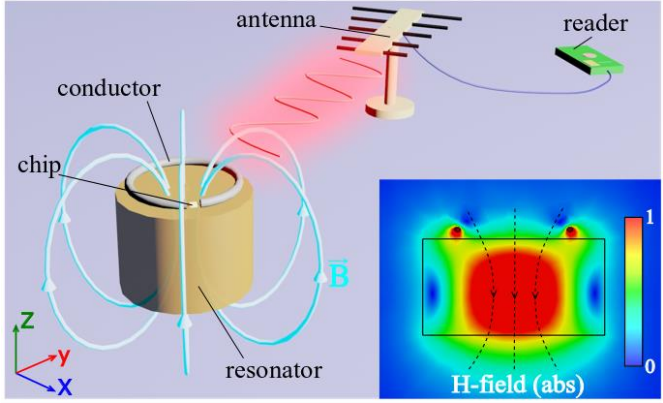

Fig. 1. Operational scheme of a miniature ceramic-based RFID tag. Displacement currents within high-Q resonator are inductively coupled to a metal split ring, functionalized with an RFID chip. Miniaturization and long reading range are ensured by localized modes within high permittivity ceramics. Inset shows the magnetic field amplitude of the fundamental magnetic mode.

The report is organized as follows: theoretical and numerical analysis for finding bounds of ceramic tags' miniaturization are discussed first. It is then followed by the experimental comparison of scattering efficiencies of new ceramic and commercially available metallic tags. Investigation of environmental fluctuation effects on tags' performances comes before experimental studies, where readout distances, achievable for different tags, are evaluated. Long-range operation with the recent state of the art equipment is evaluated by extrapolating link budget equations.

## II. SCATTERING PROPERTIES AND BANDWIDTH OF CERAMIC TAGS

Size reduction of a resonator implies the decrease of the scattering cross section bandwidth according to the Chu-Harrington limit [16]. Quality factor (Q-factor) of a structure, operating at a dipolar resonance, obeys:

$$Q > \frac{1}{ka} + \left(\frac{1}{ka}\right)^3, \quad (1)$$

where $k$ is the free space wavenumber and $a$ is the minimal radius of an imaginary sphere, inclosing the geometry. Operation with narrow bandwidth resonators can significantly reduce the capacity of a wireless communications channel. EPCGEN2 RFID standard, being the subject to standardization, requires establishing a reliable communication in 860-960 MHz frequency range, depending on a specific country. Those parameters suggest operating at Q-factors less then several thousands, which makes impossible to implement tags with sizes significantly smaller than $\lambda/100$ in a free space. Operation outside the scattering peak results in a drop of the reading distance, nevertheless a tag still can be read out.

In order to estimate the impact of the ceramic resonator miniaturization on the bandwidth, a qualitative analysis will be performed first and then followed by a more accurate theoretical investigation. The miniaturization strategy is based on a compensation of size reduction by increasing the permittivity. The resonant frequency of the fundamental magnetic $TE_{010}$ mode in the closed cylindrical resonator is given by [16]:

$$(f_r)_{npq}^{TE} = \frac{c}{2\pi a\sqrt{\varepsilon\mu}}\sqrt{x'^2_{np} + \left(\frac{q\pi a}{d}\right)^2}, \quad (2)$$

where $a$ is the radius, $d$ its height, $q$ is the axial order of the mode, $x'_{np}$ it is the $p$-th roots of the Bessel function derivative $J'_n(x'_{np}) = 0$, $\varepsilon$ and $\mu$ are the permittivity and permeability of the cylinder, respectively. We focus at the fundamental magnetic dipole mode of the resonator, thus, $q = 0$, $n = 0$ and $p = 1$, and $x'_{01}=3.83$. Equation (2) assumes PEC boundaries, which is quite an accurate approximation in the case of high-permittivity material, filling the interior of the resonator. Equation (2) shows that the permittivity of a nonmagnetic material should grow quadratically with the scaling parameter, which predicts well the numerical results, obtained with full-wave simulations. In order to estimate the dependence of the quality factor on $\varepsilon$, equations (1) and (2) should be considered. It can be shown that Q grows proportionally to a superposition of $\varepsilon^{0.5}$ and $\varepsilon^{1.5}$, giving an approximately linear dependency (on average) in the range of the considered size parameters ($ka$~0.5) (bandwidth is proportional to Q$^{-1}$). The induced current in the ring is proportional to Q, since the magnetic field flux scales linearly with later considering that the ring's area is kept constant.

While the beforehand qualitative analysis predicts the general behavior of the system's parameters, it does not provide any quantitative measure. In order to accomplish this task, electromagnetic fields, leaking from the resonator, should be calculated. While the problem of electromagnetic scattering on a finite size cylindrical resonator does not have a closed-form expression, the scenario can be well approximated with an equivalent sphere, for which Mie solutions are well known [17]. The high-permittivity dielectric sphere supports a fundamental magnetic mode, which possesses a behavior, similar to the one in the cylinder. However, Mie solutions provide the full information on the fields outside the sphere – those fields will be used to obtain currents, induced in the metal ring, assuming that the latter is weakly coupled to the ceramic resonator and only slightly changes its electromagnetic response. The current in the ring is induced by the magnetic flux of a self-consistent electromagnetic field, according to Faraday's law. In order to find the fields, the cylinder will be approximated with an equivalent sphere.

Resonances of cylinders can be found numerically by either considering the resonant structure of scattering spectra or by employing an eigenmode solver. A typical example, demonstrating the excitation of the fundamental $TE_{010}$ mode in the resonator with height $h = 17$ mm, radius $R = 20$ mm and dielectric constant $\varepsilon = 100$ appears in the inset in Fig. 1. The metal ring's radius is 11 mm. The incident plane wave is polarized along the $x$-axis, with $z$ being the symmetry axis of the cylinder. The absolute value of magnetic field distribution at the cut plane through the cylinder's center was calculated with the help of CST microwave studio, frequency domain solver. This analysis also shows the impact of a metal ring, placed on top of the ceramic resonator – the mode profile resembles the field distribution in a standalone structure, while the leaking fields excite currents in the metal ring, which acts as a perturbation. An efficient control over this perturbation with a switchable impedance of an RFID chip, placed in the gap, allows obtaining high differential RCS, which is the key for long-range operation.

The resonant frequency of the entire structure including the cylinder, metal ring, and shorten circuit was found to be $f = 900$ MHz (in fact, the parameters were chosen to hit this frequency, which falls at the middle of EPCGEN2 band). An equivalent sphere, resembling these properties, is constructed as follows: the same permittivity $\varepsilon$ is used, while the radius of the sphere is varied to match the resonant frequency of the fundamental magnetic dipolar mode. The expression for the resonant condition has a closed-form analytical form, based on Mie series. Once the equivalent radius is found, the fields outside the sphere can be calculated as:

$$\vec{H}_S = \sum_{n=1}^{\infty} i^n \frac{2n+1}{n(n+1)} E_0 \left[ib_n\vec{N}_{o1n}^{(3)} + a_n\vec{M}_{e1n}^{(3)}\right], \quad (3)$$

where $\vec{N}_{o1n}^{(3)}$ and $\vec{M}_{e1n}^{(3)}$ are the vector spherical harmonics, $E_0$ the amplitude of an incident plane wave [15], $a_n$ and $b_n$ are the Mie coefficients.

The fundamental magnetic mode corresponds to a single nonvanishing coefficient in the series, namely $b_1$. The total magnetic flux through the ring is calculated by integrating the total field. The nontrivial part of the magnetic flux, corresponding to the scattered field, is calculated by integrating (3):

$$\Phi_{sc} = \frac{3\pi a E_0}{k_0 Z_0} b_1 h_1^{(1)}(k_0 a)(\alpha + 0.5\sin(2\alpha)), \quad (4)$$

where $a$ is the sphere radius, $k_0$ is the wave number, $Z_0$ is the impedance of free space, $h_1^{(1)}(k_0 a)$ is the spherical Hankel function of the first kind and $\alpha = arctan(r/a)$. Then current in the ring is calculated as:

$$I(\omega) = \Phi_{sc}/\sqrt{L^2 + (R_\Omega/\omega)^2}, \quad (5)$$

where $R_\Omega$ is resistance and L is inductance of the ring. In our case $L = 4.38 \cdot 10^{-8}$ H, which is several orders higher than $R_\Omega/\omega$ and hence resistance can be neglected. It is worth noting that the self-inductance of the ring can be introduced self-consistently in the model, but this complication does not provide any significant advantage to the basic understanding of the process. Equation (4), plotted as a function of frequency allows calculating the bandwidth, which is defined as the full width at half maximum of $I(\omega)$ spectrum.

At the next step, theoretical predictions are assessed versus full-wave numerical analysis. For this purpose, several cylindrical resonators are studied and the impact of miniaturization on bandwidth and current are investigated. The largest considered structure has the following parameters: $R = 19$ mm, $h = 22$ mm, $\varepsilon = 80$, and the resonant frequency is 900 MHz – the last parameter is kept constant through all the designs. The size of this initial resonator is then reduced by a multiplicative scaling factor. In order to keep the resonant frequency constant, the value of ε was adjusted with the help of the numerical routine.

Figure 2 shows the operational bandwidth (panel a) and induced current in the metal ring (panel b). The ring's radius is the same (11 mm) in all the realizations. The value of the current indicates whenever the electronic chip, which will be plugged within the ring's gap, will operate properly. Typical threshold values for the tag 'wake up' are in the range of μW, depending on an integrated circuit (IC) architecture. Operational bandwidth corresponds to the wireless channel capacity, through which the data will be transformed. Both of those values are crucial for a proper RFID tag functioning.

Numerical and theoretical models predict that the resonator's size reduction causes the drop in the bandwidth (in accordance with the Chu-Harrington limit), while the current grows with the increase of the quality factor, as it was also predicted by the qualitative analysis. It is quite remarkable that the theoretical approximation (4) is in a very good correspondence with the numerical analysis. Equation (4) can be used to estimate the minimal size of a ceramic tag, suitable for a reliable readout. Since this link budget calculation strongly depends on the reader's and IC's parameters and specifications of a probable application, this step will be skipped here. The formula can be straightforwardly applied on any practical scenario.

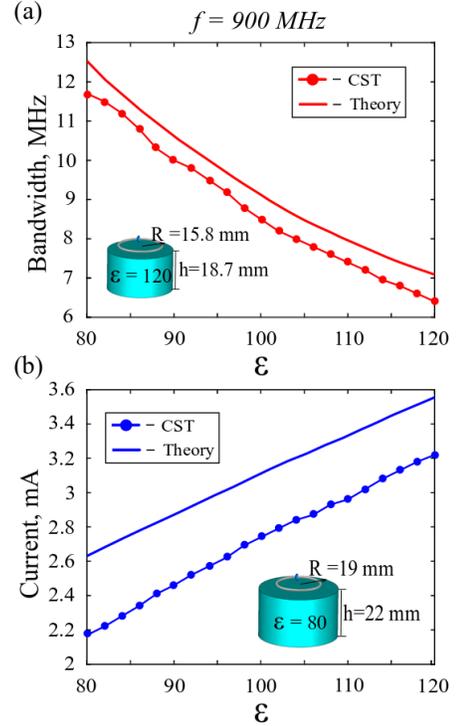

Fig. 2. Electric performance of ceramic tags as the function of their permittivity. Aspect ratio between the radius and height is constant ($R/h = 0.85$), the resonant frequency is $f_0 = 900$ MHz. Absolute values of $R$ and $h$ are uniquely defined by $\varepsilon$ in order to keep the resonance frequency constant. (a) Current's bandwidth. (b) Conduction current in a ring, placed on top or the resonator. Radius of the ring is 11 mm in all of the realizations (the gap is shorted). Solid lines – theoretical prediction, based on (4). Solid dotted line is the full-wave numerical analysis.

## III. ELECTROMAGNETIC PROPERTIES OF CERAMIC RFID TAGS

While near-field coupling governs the performance of the RFID readout in standard applications, an increase of the interrogation distance forces the system to operate at far-field conditions. Hereinafter the readout range will be above several meters, prevailing Fraunhofer distances by an order of magnitude and, hence, ensuring far-field conditions.

The parameters of the fabricated ceramic resonators are shown in Table 1 and were adjusted to make the structures resonating at UHF RFID frequencies (860- 920 MHz). A foil split ring with radius 11 mm was placed on top of the resonators and an RFID chip (Impinj Monza 4), detached from a commercial tag, was soldered within the gap. One of the crucial parameters is the matching between the IC and the antenna (ceramic cylinder, in this case). Conjugated impedance condition should be approached in order to maximize differential RCS.

**Table I**. Parameters of the fabricated ceramic resonators.

| Resonator | $R$ | $h$ | $\varepsilon$ |
|---|---|---|---|
| 1 | 19 mm | 22 mm | 80 |
| 2 | 17 mm | 20 mm | 100 |

To demonstrate and optimize the matching conditions, the following numerical analysis has been performed. Numerical port with $Z = 4.9 - i70$ ohm impedance was plugged within the ring's gap and complex reflection ($S_{11}$ parameter) spectra were calculated. The impedance value at the resonance corresponds to the data sheet of the IC. Figure 3 shows the spectra and underlines an existence of a relatively good matching at the resonance frequency. While the IC's impedance, taken from the datasheet, it is frequency dependent, its dispersion is rather minor in comparison to the sharp peaked resonance and, hence, it can be neglected. It is worth noting that no additional lumped elements were used to perform the matching, which is achieved via mutual inductive coupling between the resonator and the ring. The matching can be further improved by a slight modification of the layout, e.g. the distance between the surface of the cylinder and the foil ring of a geometry of the later.

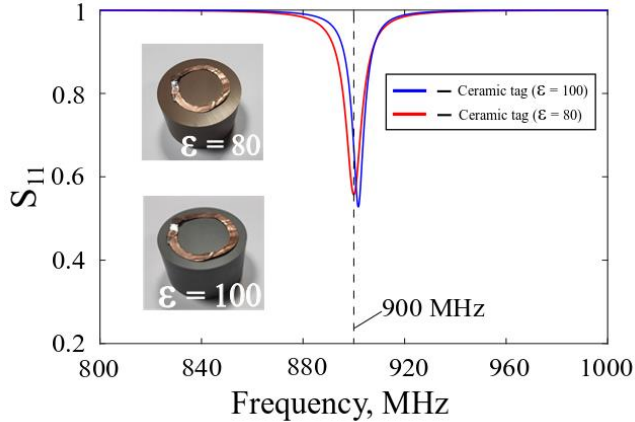

Fig. 3. Numerically calculated $S_{11}$ spectra of ceramic tags, indicated in legends. $S_{11}$ parameters were extracted by plugging a port directly to the ring's gap. The impedance of the port is equal to the value, provided by the vendor IC (Impinj Monza 4). Insets shows photographs of the fabricated tags.

## IV. ROBUSTNESS TO TEMPERATURE FLUCTUATIONS

Another important requirement for a proper RFID operation is the tolerance to environmental fluctuations. Those might include extreme temperature, moisture, chemical ambient, and any other hash condition. Ceramic composites have significant advantages with respect to metal structures in this context. Here we will check the influence of the temperature change on tag's operation, which might be quite substantial in the case when resonant structures are in use.

While there is no data on temperature-dependent permittivity for specific ceramics, which is in use here, related structures have been investigated [18], [19]. An approximate temperature-induced differential drift in the permittivity is $\Delta\varepsilon/\Delta T = 0.01/°C$, which is quite minor. Figure 4 shows the effect of the permittivity fluctuation on the RCS and on the current, which drives the tag. The results were obtained numerically. A cylinder with fixed parameters ($R = 17$ mm, $h = 20$ mm, thermal expansion is neglected) was illuminated with a plane wave with an amplitude of 1 V/m. The relative permittivity was scanned in the range between 80 and 120. Resonant frequency of 900 MHz corresponds to $\varepsilon = 100$, which is exactly at the center of the considered range. Values of the current were calculated for the case where a small resistor (0.1 Ohm) is plugged into PEC ring with the radius of 11 mm.

The results suggest in Fig. 4 that the permittivity fluctuations on the order of unity (full width at half maximum) will still allow performing the readout. This range corresponds to 100-200 °C temperature change, which is more than sufficient for the normal operation. Standard tags are declared to operate in the interval between -20 and 80 °C.

It worth briefly noting that our architecture still contains unshielded metal elements (the ring and the chip). However, those parts can be embedded in the interior of the ceramic resonator, which will shield them from an environment to a certain extent.

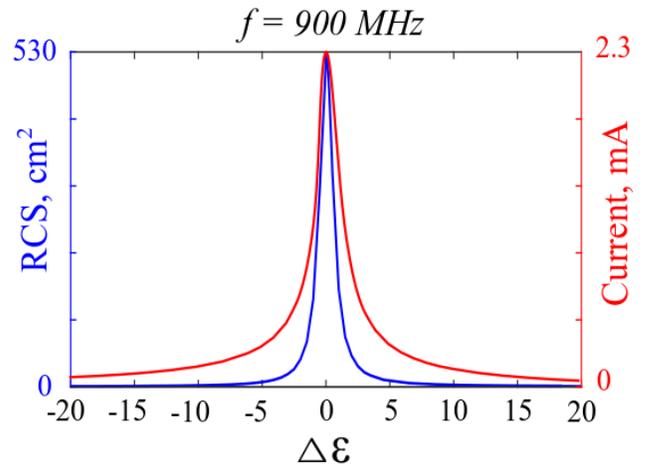

Fig. 4. Ceramic tag parameters as the function of the permittivity detuning. Left scale (blue) – RCS, right scale (red) – the current trough a small resistor. Additional parameters are given in the main text.

## V. EXPERIMENTAL DEMONSTRATION OF THE READING RANGE: CERAMIC VERSUS STANDARD

In order to assess the performances of new tags, a comparative experiment has been performed. Two ceramic resonators and three commercial UHF metallic tags [thin electric dipoles, see inset in Fig. 5(b)] Alien 9662, Impinj HR61 and Impinj E42 were evaluated in long-range readout configuration (Fig. 5(a) is the photograph of the experiment layout). A reader device (model KLM900S) was plugged into a laptop via USB port and the amplitude of the received signal was monitored via the software, provided by the vendor. This device comes with a standard PCB patch antenna, which is not designed to provide long-range reading distances due to its low gain. Hence, this element was replaced with a custom-made Yagi-Uda antenna (4 directors), with -15 dB matching in 880-960 MHz frequency range inset [Fig. 5(a)]. The gain of 9.3 dBi has been estimated. All five tags were placed (one after another) at the antenna E-plane and accurately moved away from the minimal distance of 50 cm to the maximal of 5 m (Fig. 5(a) for the setup layout). The received signal intensity was monitored for each physical position (time average has been performed to reduce fluctuations). The noise level of the reader (the minimal detectable intensity) is -50 dBm. Figure 5(b) demonstrates the received signal power as the function of the distance between the tag and the reader. The maximal reading distance for commercial metallic elements in the current arrangement is less than 4 m (after further increase of the distance the signal goes below -50 dBm and becomes non-processable). The ceramic tags are readable from distances, approaching 5 m.

In order to underline the main mechanism for the readout distance improvement, the link budget of the system has been evaluated. There are two major parameters, related to the hardware implementation, have to be verified. The first factor is the chip activation threshold. RFID tag has a rectifier, which powers the IC. A certain threshold intensity should be obtained on the tag's antenna port to power the electronics. The second factor is a reader's sensitivity. The intensity of the backscattered signal should be within the dynamical range of the receiver. The beforehand mentioned factors should be assessed simultaneously, since the weakest part of the system will limit the range.

The maximal reading distance is given by the Friis' uplink model [20]–[22]:

$$L = \frac{\lambda}{4\pi} \sqrt{\frac{P_t G_{TR} G_t \tau}{P_{ch}}}, \quad (6)$$

where the parameters are given in Table 2.

**Table 2.** Parameters of the RFID system for link budget calculations.

| Parameter | Physical meaning | Value |
|---|---|---|
| $P_t$ | Power, transmitted by the reader | 0.1 W |
| $G_{TR}$ | Gain of the reader transmitter/receiver antenna | 8.5 dB |
| $G_t$ | Gain of the ceramic tag | 1.44 dB (ε =80) <br> 1.51 dB (ε =100) |
| $P_{ch}$ | Chip sensitivity | $1.8 \cdot 10^{-5}$ W |
| $P_r$ | Reader sensitivity | $10^{-8}$ W |
| $\lambda$ | Wavelength | 0.33 m |
| $\tau$ | Power transmission coefficient | 0.51 (ε =80) <br> 0.56 (ε =100) |
| ME | Modulation efficiency | 0.7 |

Few of the parameters in Table 2 require a special attention. The power transmission coefficient ($\tau$) determines the matching between the tag antenna and the chip and this value ranges as $0 \leq \tau \leq 1$. Here we calculated this factor numerically by extracting the data from Fig. 3 ($1 - |S_{11}|$). Values in Table 2 suggest a room for an additional improvement. The gain of ceramic tags resembles the one of a dipolar resonator, which is quite expected. The values were estimated with a standard gain calculation routine, implemented in CST Microwave Studio.

The sensitivity of the basic low-cost IC (Impinj Monza 4) is -17 dBm, which corresponds to the activation range L=5.3 m (Eq. 6). The improved version (Impinj Monza 6 generation) already has the sensitivity of -22 dBm, which will allow the activation at a distance L=9 m. Using new generation of sensitive electronic components allows improving the reading distances quite significantly.

The second factor, which might limit the range, is the reader sensitivity. In this case the tag is assumed to be activated, and the differential RCS plays the key role. In this case, the modulation efficiency (ME) parameter is defined. Taking into account the readers minimal received power ($P_r$), the monostatic radar equation reads as [23]:

$$L' = \frac{\lambda}{4\pi} \sqrt[4]{\frac{P_t G_t^2 G_{TR}^2 ME}{P_r}}, \quad (7)$$

In practice, ME ranges in $0 \leq ME \leq 4$ or $ME \leq 6\ dB$ [23]. In our case, $ME \approx 0.7$. This value was estimated numerically by calculating backscattering of the tag in two cases, which correspond to different nominals of lumped impedances, taken from Impinj Monza 4 data sheet.

Reader device with a moderately low dynamic range can activate a tag at a distance but cannot process the received signal. In this case, the sensitivity is the bottleneck. In our investigation we employed a relatively cheap device, which cannot read signals below -50 dBm (also can be seen in Fig. 5). State of the art readers (e.g. Alien ALR-9900) can interpret received signals at -80 dBm level, which significantly improves the reading range.

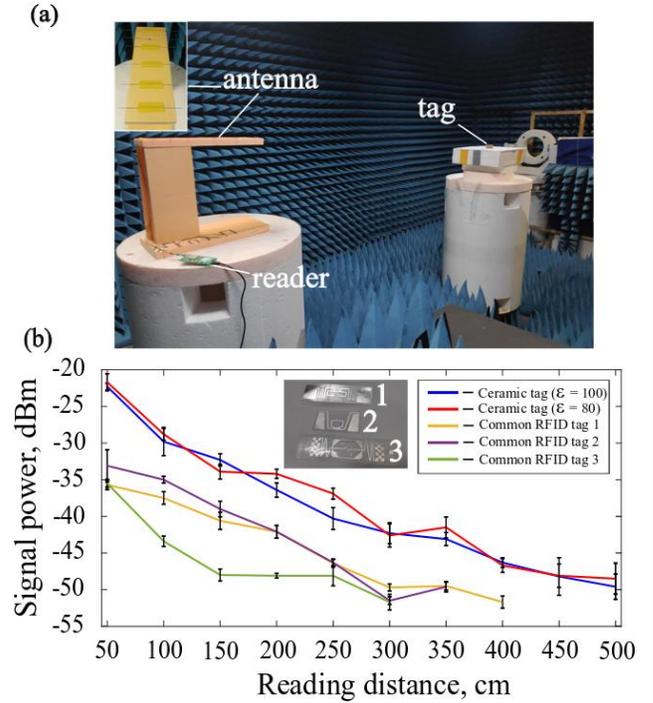

Fig. 5. (a) Photo of the experimental setup in an anechoic chamber – RFID reader with a custom-made Yagi-Uda antenna. (b) Received signal power as the function of the distance between the reader and the tag. Inset shows photograph of the measured commercial tags.

Finally, the link budget equations can allow extrapolating the reading distance in the case, when state of the art reader and tag's IC are used. For example, -22 dBm chip's sensitivity allows obtaining around 23 m reding distance with our ceramic-based architecture (Eq. 6). Improving the reader (-80 dBm sensitivity is commercially available) and its Equivalent Isotopically Radiated Power (EIRP) to 4 W (further increase is restricted by international regulations) will allow reading an activated tag from a distance of 41 m (Eq. 7). In overall, the bottleneck in this scenario is the IC and, hence, the extrapolated readout distance is 23 m. It is worth noting, that the existent demonstrated reading distance is about 26 m, reported for a rather complex cumbersome tag with a high gain and improved matching circuitry [24].

## VI. CONCLUSION

The new type of RFID tags, based on ceramic resonators, has been investigated and aimed at achieving miniaturization and extension of readout distances. Apart from the applied aspect of the investigation, a new physical concept in the field has been introduced. In particular, it has been shown that displacement currents, resonantly exited within the resonator's volume, are efficiently coupled to conduction current in a small loop, functionalized with an RFID chip. This approach allows for designing robust and environmentally sable ceramic components, which drive the entire process. From the applied standpoint, this new architecture provides new routes for achieving small footprint RFID tags, which support long reading range, which is valuable in numerous applications, such as retail, security, IoT, and many others.